\documentstyle[epsf,epsfig,axodraw,multicol,aps]{revtex}
\def\ruleleft{\vspace{-2.5\baselineskip}\begin{multicols}{2}\ \linebreak\vspace{-\baselineskip}\hrulefill\raisebox
{0.84mm}{$\!\rfloor$}\[\]\end{multicols}\vspace{-1.5\baselineskip}}
\def\ruleright{\vspace{-1.5\baselineskip}\begin{multicols}{2}\ \linebreak\raisebox
{-2.45mm}{$\lceil\!$}\hrulefill\end{multicols}\vspace{-\baselineskip}}
\def\axowidth{1.0 }
\def\axoscale{0.6 }
\begin{document}
\title{\bf When Do Like Charges Attract?}
\author{\bf Yan Levin}
\address{\it Instituto de F\'{\i}sica, Universidade Federal
do Rio Grande do Sul\\ Caixa Postal 15051, CEP 91501-970, 
Porto Alegre, RS, Brazil\\
{\small levin@if.ufrgs.br}}
\maketitle
\begin{abstract}
We study the interaction potential between two polyions inside a
colloidal suspension.  It is shown that at large separation the interaction
potential is purely repulsive, with the induced attractive interactions
being doubly screened.  For short separations
the condensed counterions become correlated, what leads to an effective attraction between the two
macromolecules.   
\end{abstract}


\bigskip

\centerline{{\bf PACS numbers:} 05.70.Ce; 61.20.Qg; 61.25.Hq}

\begin{multicols}{2}

The systems in which the interactions between the 
particles are predominantly due to the long-ranged
Coulomb force remain an outstanding challenge to 
classical statistical 
mechanics.  Even such basic question as the existence
of a phase transition in a symmetric electrolyte has 
remained uncertain until quite recently \cite{st76}.  Although the 
answer to this question has proven to be affirmative \cite{fis93}, 
the true nature of the transition as well as its 
universality class still remain unclear \cite{fis96}.  Surprisingly the
theory that is in closest agreement with the Monte
Carlo simulations is based on the old ideas first introduced
by Debye, H\"uckel, and Bjerrum \cite{dh23} more than seventy
years ago.  The fundamental insight
of Debye was to realize that since the mean force 
inside the electrolyte is zero, it is the correlations
in positions of oppositely charged ions that produce the main 
contribution to the free energy.  A great advantage
of the Debye-H\"uckel ($DH$) theory, besides its simplicity, is that its
 linear structure allows it
to avoid the internal inconsistencies that are often present in
the more complicated non-linear theories of electrolytes \cite{on33}.  
The price for linearity,
however, is that the $DH$ theory cannot account for the non-linear configurations,
such as formation of dipoles, which become important at low temperatures.  
It was an idea of Bjerrum, proposed only
three years after the publication of original $DH$ paper, that the missing
non-linearities can be reintroduced into the $DH$ theory through the assumption
of chemical equilibrium between the monopoles and multipoles.  The
extended Debye-H\"uckel-Bjerrum ($DHBj$) 
theory has proven extremely successful in elucidating the
underlying physics of symmetric electrolytes \cite{fis96}, polyampholytes \cite{lev95}, rod-like polyelectrolytes \cite{lev96}, and charged colloidal suspensions \cite{lev98}, 
with its validity extending
far into the regime where the pure linearized $DH$ theory fails. 
In this letter we shall use the $DHBj$ theory to 
explore one of the most fascinating phenomena in condensed matter physics, the appearance of attraction between two like-charged polyions inside a colloidal suspension \cite{ise84}.

To model a charged colloidal suspension we use a restricted primitive model, in 
which the polyions are treated as hard spheres of radius $a$ and uniform 
surface charge, $\sigma_-=-Zq/4\pi a^2$, and the counterions are point
particles of charge $zq$.  The solvent is modeled as a uniform medium of
dielectric constant $D$. In Ref. \cite{lev98} it was demonstrated
that the equilibrium state of the
colloidal suspension consists of some free polyions of density
$\rho_0$, free counterions of density $\rho_f$, and of clusters composed of
{\it one } polyion and of ${1 \le n \le Z/z}$ associated counterions.  
The effective charge of a $n$-cluster is
$Z_{\rm eff}=-(Z-nz)q \equiv 4 \pi a^2 \Delta \sigma$. The $DHBj$ theory
allows us to explicitly calculate the distribution of cluster densities \{$\rho_n$\} \cite{lev98}.

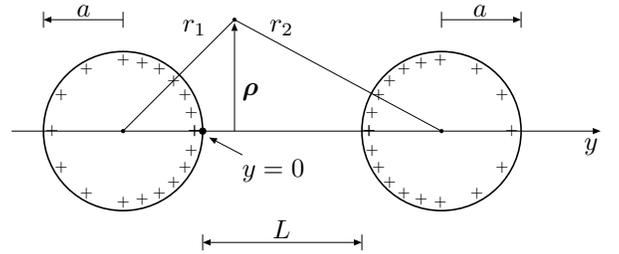
\begin{figure}[hb]
\begin{center}
\begin{picture}(400,200)(-90,-50)
\BCirc(0,0){50}
\BCirc(200,0){50}
\def\axowidth{0.5 }
\Line(0,0)(70,70)
\Line(70,70)(200,0)
\LongArrow(-70,0)(300,0)
\LongArrow(100,-70)(50,-70)
\LongArrow(100,-70)(150,-70)
\Line(50,-75)(50,-65)
\Line(150,-75)(150,-65)
\LongArrow(0,70)(-50,70)
\LongArrow(200,70)(250,70)
\Line(0,65)(0,75)
\Line(-50,75)(-50,65)
\Line(200,65)(200,75)
\Line(250,75)(250,65)
\GCirc(50,0){2}{0}
\LongArrow(70,0)(70,68)
\Text(75,25)[l]{\boldmath$\rho$}
\LongArrow(75,-15)(54,-4)
\Text(75,-25)[l]{$y=0$}
\Text(295,-10)[c]{$y$}
\Text(-25,76)[c]{$a$}
\Text(225,76)[c]{$a$}
\Text(100,-62)[c]{$L$}
\Text(45,65)[c]{$r_1$}
\Text(100,65)[c]{$r_2$}
\GCirc(0,0){1}{0}
\GCirc(200,0){1}{0}
\GCirc(70,70){1}{0}
\Text(45,0)[c]{\tiny$+$}
\Text(43,12)[c]{\tiny$+$}
\Text(39,23)[c]{\tiny$+$}
\Text(32,32)[c]{\tiny$+$}
\Text(23,39)[c]{\tiny$+$}
\Text(12,43)[c]{\tiny$+$}
\Text(45,0)[c]{\tiny$+$}
\Text(43,-12)[c]{\tiny$+$}
\Text(39,-23)[c]{\tiny$+$}
\Text(32,-32)[c]{\tiny$+$}
\Text(23,-39)[c]{\tiny$+$}
\Text(12,-43)[c]{\tiny$+$}
\Text(0,45)[c]{\tiny$+$}
\Text(-23,39)[c]{\tiny$+$}
\Text(-39,23)[c]{\tiny$+$}
\Text(-45,0)[c]{\tiny$+$}
\Text(0,-45)[c]{\tiny$+$}
\Text(-23,-39)[c]{\tiny$+$}
\Text(-39,-23)[c]{\tiny$+$}
\Text(155,0)[c]{\tiny$+$}
\Text(157,12)[c]{\tiny$+$}
\Text(161,23)[c]{\tiny$+$}
\Text(168,32)[c]{\tiny$+$}
\Text(177,39)[c]{\tiny$+$}
\Text(188,43)[c]{\tiny$+$}
\Text(155,0)[c]{\tiny$+$}
\Text(157,-12)[c]{\tiny$+$}
\Text(161,-23)[c]{\tiny$+$}
\Text(168,-32)[c]{\tiny$+$}
\Text(177,-39)[c]{\tiny$+$}
\Text(188,-43)[c]{\tiny$+$}
\Text(200,45)[c]{\tiny$+$}
\Text(223,39)[c]{\tiny$+$}
\Text(239,23)[c]{\tiny$+$}
\Text(245,0)[c]{\tiny$+$}
\Text(200,-45)[c]{\tiny$+$}
\Text(223,-39)[c]{\tiny$+$}
\Text(239,-23)[c]{\tiny$+$}
\end{picture}
\end{center}
\vspace{1cm}
\begin{minipage}{0.48\textwidth}
\caption{The two clusters separated by a distance $R=2a+L$.  The polyions have a fixed surface charge density $\sigma_-$ and the mobile surface charge $\sigma_+$ due to the condensed counterions, $+$. The overall imbalance of charge, $|\sigma_-|>|\sigma_+|$ is responsible
for the polarization of clusters and for decrease in the electrostatic energy 
relative to the uniform distribution of charge, $\Delta \sigma=\sigma_+ + \sigma_-$.}
\end{minipage}
\end{figure}

Consider two clusters separated by a distance $R=2a+L$,  
each consisting of a polyion and $n$ condensed counterions. 
It is convenient to set up a bipolar coordinate system with  
the two polyions located at $r_1=0$ and $r_2=0$, respectively.  To calculate the
effective potential between the two clusters we shall appeal to the $DHBj$ theory \cite{lev98}.
The electrostatic potential inside the suspension satisfies the Poisson equation,
\begin{equation}
\label{eq1}
\nabla^2 \phi= \left\{\begin{array}{cl} 0,  \quad & \mbox{ for } r_1<a \mbox{ or } r_2<a,  \\
-4 \pi \rho_q({\bf r}_1, {\bf r}_2) /D, \quad &\mbox{ for } r_1\ge a \mbox{ and } r_2 \ge a. \end{array}\right.
\end{equation}
Within the $DHBj$ theory the charge density is approximated by 
\begin{eqnarray}
\label{eqn2} 
\rho_q({\bf r}_1,{\bf r}_2) &=&-\sum_{n=0}^{Z/z}(Z-zn) q \rho_n +
z q \rho_f  e^{-\beta z q \phi({\bf r}_1,{\bf r}_2)}\nonumber \\
&+&\sigma_-
 \delta(r_1-a)+\sigma_+ e^{-\beta z q \phi({\bf r}_1,{\bf r}_2)} \delta(r_1-a)
\\
&+& \sigma_- \delta(r_2-a)  + \sigma_+e^{-\beta z q \phi({\bf r}_1,{\bf r}_2)} \delta(r_2-a), \nonumber
\end{eqnarray}
where, $\beta=1/k_B T$, and the average surface charge density of 
condensed counterions is $\sigma_+ = nzq/4 \pi a^2$.
Note that only the counterions are assumed to get polarized, while the polyions
and clusters contribute only to the neutralizing background.  
As a second approximation, {\it linearization} leads to 
$\rho_q({\bf r}_1,{\bf r}_2)=-(D/4 \pi)  \kappa^2 \phi({\bf r}_1,{\bf r}_2)  
 + [\Delta \sigma - D \alpha \phi({\bf r}_1,{\bf r}_2)/2\pi] [\delta(r_1-a) + \delta(r_2-a)]$, where the inverse  Debye and the Gouy-Chapman lengths are
respectively $\kappa= \sqrt {4 \pi \beta z^2 q^2 \rho_f}$, and 
 $\alpha=2\pi \sigma_+ \beta zq/D.$  Naively one
might think that linearization is valid only at high temperatures.  This,
however, is not the case, since the non-linearities are effectively included in 
the renormalization of the polyion charge by the 
formation of clusters \cite{lev98,alex84}. 
The set of equations (\ref{eq1})  is extremely difficult to study due to
non-trivial boundary conditions.  Nevertheless some progress can be made
in the two limiting cases, $L \gg a$ and $L \ll a$.  All the detail of calculations
will be presented elsewhere,  here we shall content ourselves with giving a 
simple and an intuitive explanation of the results.  
	
In the limit $ L\gg a$ and $ \kappa L >1$
one finds that the crucial small parameter
 is $\epsilon=\exp(-\kappa R)/R$ \cite{li94}.  
It is then
possible to show that the interaction potential between the 
two $n$-clusters is
\begin{equation}
\label{eqn3}
W(R)=Z_{\rm eff}^2 \theta^2(\kappa a) \frac{e^{-\kappa R}}{D R}- 
Z_{\rm eff}^2  \kappa^2 a^3 \theta^2(\kappa a) h(\alpha a) 
\frac{e^{-2 \kappa R}}{D R^2} ,
\end{equation}
where the two scaling functions are $\theta(x)=e^{x}/(1+x)$ and 
$h(x)=\frac{2}{3}-\frac{3}{2 x} \: ln(1+\frac{2 x}{3})$. 
The first term is just the usual DLVO potential between the two colloidal 
particles with the effective charge $Z_{\rm eff}$ \cite{der41}.  The second term is the 
result of two types of interactions. First, presence of polyions
produces  holes in the ionic atmosphere.  Since the charges of the holes
have the same sign as the charges of the polyions, the holes interact repulsively
with the polyions \cite{li94}.  The second effect is that the mobile counterions confined
to the surface are easily polarized.  Thus, the electric field produced by the 
cluster ${\cal C}_1$ induces a dipole moment in the cluster ${\cal C}_2$, and vice-versa.  
The  polarizability of a $n$-cluster is found to be 
$\gamma=2\alpha a^3/(3+2\alpha a)$.  The effective dipole moment interacts attractively with the charge that induced it.   Since the
scaling function $h(x)$ changes sign, we see that the correction
to the DLVO potential is repulsive (hole dominated) for $\alpha a < 1.716...$ and is
attractive (dipole dominated) for $\alpha a > 1.716...$.  However, we do note that
this induced attraction is  ``doubly'' screened and is dominated by the leading
order DLVO term.  
	
We shall now turn our attention to the opposite limit $ L\ll a$ and $ \kappa L <1$.
Under these conditions the relevant small parameter is $ \epsilon= L/a$, 
and the Debye screening due to unassociated counterions can be neglected.   
It is convenient to define a cylindrical coordinate
system, $(\mbox{\boldmath$\rho$}, y)$, 
with the $y$ axis passing through the centers of
the two clusters, Fig 1.  The two polyions are located at $y=-a$ and $y=a+L$,  respectively.  The
interaction potential can be subdivided into two parts.  The mean-field repulsion, $V_{MF}$, arising from the overall imbalance of charge on the two clusters, and the attraction coming from the correlations among the condensed counterions.  
It is evident that the repulsive energy satisfies, $V_{MF} \le Z_{\rm eff}^2/ D R$.  In what follows we shall approximate $V_{MF}$ by the 
upper bound, $V_{MF}= Z_{eff}^2/ D R$,  that is, we shall treat the mobile charge, $\sigma_+$, as, on average,  
{\em uniformly  distributed} on the surface of the two polyions.   Clearly this
assumption will {\it overestimate} \ the importance of repulsion, since it neglects
the ability of the bound counterions to arrange themselves in the most efficient way
to minimize the energy, see Fig. 1.  With this in mind, to leading order in $\epsilon$, the geometry can be replaced by that of two plates located at $y=0$ and $y=L$, each with a fixed surface
charge, $\sigma_-$, and the mobile charge, $\sigma_+$.  The free energy can
be calculated using the $DH$ theory.  Let us fix one counterion at $({\bf 0},0)$.  
The equations (\ref{eq1}) may now be integrated to  give
\begin{equation}
\label{eqn4}
\phi(\rho, y)=\frac{\pi\Delta\sigma}{D}L - \frac{2\pi\Delta\sigma}{D}(|y|+|y-L|)+
\Delta \phi_c(\rho, y),
\end{equation}
where the correlation potential can be written in terms of the 
zero-order Bessel function,
\end{multicols}
\ruleleft
\medskip

\begin{eqnarray}
\label{eq5}
\Delta \phi_c(\rho,y)=  \left\{\begin{array}{cl} \int_{0}^{\infty} A(k) e^{ky} J_0(k \rho) dk , \quad  \quad & \mbox{ for } y\le  0 , \\
\int_{0}^{\infty} [B(k) e^{ky}+ C(k)e^{-ky}]  J_0(k \rho) dk ,
\quad  \quad & \mbox{ for }  0 < y <  L ,\\
 \int_{0}^{\infty} E(k) e^{-ky} J_0(k \rho) dk , \quad  \quad & \mbox{ for } y \ge  L . \end{array}\right. 
\end{eqnarray}

\ruleright
\begin{multicols}{2}
\noindent{}The coefficients, $A,B,C,E$, are determined through the conditions
of continuity of the potential and discontinuity of the normal component of the electric field related to
the presence of surface charge at $y=0$ and $y=L$,  
 $\sigma(\rho,0)=\Delta\sigma - \sigma_+  z q \beta\phi(\rho,0) + z q\delta(\rho)/
2\pi \rho$ and  
$\sigma(\rho,L)=\Delta\sigma-\sigma_+ z q \beta\phi(\rho,L)$, respectively.
For our purpose
it is sufficient to determine only the coefficient $A(k)$,  since this is the only one
that will enter into the calculation of free energy. We find
$A(k)=[k(k+\alpha)-k \alpha e^{-2kL}]/[(k+\alpha)^2- \alpha^2 e^{-2kL}]$.  
The electrostatic potential of the counterion due to the presence of other charges
is   $\lim_{\rho \to 0}(\phi(\rho, 0)-zq/D\rho)$ \cite{eso97}.   The regularized free energy 
per unit area is then obtained through the Debye charging process, in which all the particles 
in the system are charged from $0$ to their final charge,
\end{multicols}
\ruleleft
\medskip

\begin{equation}
\label{eq6}
f_R=-\frac {2 \pi (\Delta\sigma)^2}{D} L\int_{0}^{1}
\lambda d \lambda + \frac{2 z q \sigma_+}{D} \int_{0}^{1} \lambda d\lambda  \int_{0}^{\infty}
\left[\frac {(k+\lambda^2\alpha)-\lambda^2 
\alpha e^{-2kL}}{(k+\lambda^2\alpha)^2- \lambda^4\alpha^2 e^{-2kL}}
-\frac{1}{k+\lambda^2\alpha} \right] kdk
\end{equation}

\ruleright
\begin{multicols}{2}

The force per unit area that each polyion exerts on the other is 
$F=-\partial f_R/\partial L$.  Performing the differentiation we find
$F=\frac{\pi (\Delta\sigma)^2}{D} + \frac{g(\alpha L)}{\beta L^3}$, with the scaling function $g(x)$ plotted in Fig 2.  The first
term is the repulsion due to  excess charge on the two clusters, while the
second term is the correlation-induced attraction \cite{att88}.
 For $x \gg 1$, $g(\infty)=-\zeta(3)/8\pi$, where $\zeta$ is the Riemann
zeta function.  For small $x$, 
$g(x) \approx -x^2/4 \pi$, and we note that for short enough separation the correlation-induced
 attraction will always dominate the excess charge repulsion!  Furthermore, as was mentioned earlier,
taking a better account of the spherical geometry
will only favor the attraction by decreasing the repulsive contribution, $V_{MF}$, and increasing the correlations ($\alpha$) between the induced counterions on the two hemispheres facing each other, Fig 1.
Let us now make some estimates.  Suppose that the condensed
 counterions neutralize a
fraction of fixed charge, $|\sigma_+|=f |\sigma_-|$.  We shall see attraction
when $\mu \equiv z^2 \lambda_B f^2/(1-f)^2 > L$, where the Bjerrum length,
 $\lambda_B= \beta q^2/D$, is $7.2 $ \AA\  in water.  Consider a colloidal 
suspension of polystyrene
particles with characteristic size 700 \AA \ and charge $Z=1000$.  
In Ref. \cite{lev98} it was
found that the average cluster size is $<n>=400$ or $f=0.4$, what leads to $\mu=3.2 $ \AA. 
This is smaller than the size of a hydrated counterion.  We, therefore,
do not expect that for colloids with $Z<1000$ one will observe any
attraction. However, for polyions with $Z=3000$, $f=0.73$, the attractive interaction will dominate for 
$L<53 $ \AA \ and may be observable experimentally.  If a 
multivalent salt is added to the colloidal suspension, due to a strong 
electrostatic
attraction the multivalent counterions will be preferentially 
adsorbed to the polyion surface, leading to
even stronger fluctuation-induced attraction.  We like to emphasize that a presence of attraction does not imply the existance of a phase transition.  The thermodynamic properties of a colloidal suspension are mostly determined by the counterions and their interaction with the polyions and clusters.  The contribution of the cluster-cluster interaction to the osmotic pressure is minimal \cite{lev98}. Nevertherless the metastable effects connected with the presence of attractive forces might explain the unusual observations connected with the charged colloidal suspensions \cite{ise84}.  Finally, we should stress that
the attraction predicted by the $DHBj$ theory is intrinsically a finite-concentration
result and will disappear at infinite dilution.  The reason for this is that, in the
case of spherical colloids, the formation of clusters is a purely finite-concentration
phenomena.  This should be contrasted
 with the rod-like polyelectrolytes, for which the strong logarithmic potential existing between the polyions and the counterions allows for the
condensed layer to persist all the way to zero density \cite{lev96}, making it viable to study 
just {\it two} polyions with their counterions \cite{oo68}.

\begin{figure}[hb]
\begin{center}
\leavevmode
\epsfxsize=0.48\textwidth
\epsfbox[0 20 580 450]{
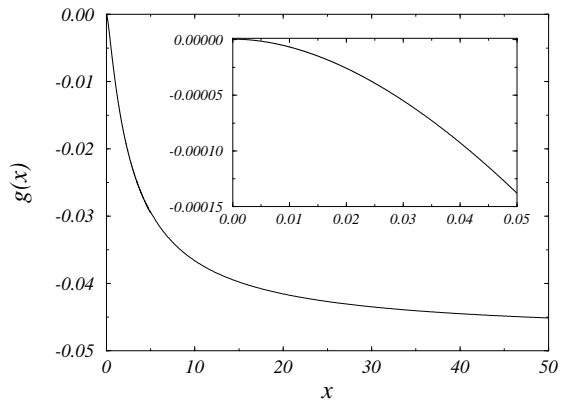}
\end{center}
\begin{minipage}{0.48\textwidth}
\caption{The scaling function $g(x)$.}
\end{minipage}
\end{figure}

The author is grateful to Dr. M. N. Tamashiro for carefully checking
the calculations.  It is a pleasure to thank Prof. P. Pincus for interesting
conversations and for sending preprints of his work prior to publication. This 
work was supported by CNPq and CAPES, Brazilian science agencies.

\end{multicols}

\end{document}